\documentclass[%
 reprint,
 amsmath,amssymb,
 aps,
pra,
]{revtex4-1}

\usepackage{graphicx}% Include figure files
\usepackage{dcolumn}% Align table columns on decimal point
\usepackage{bm}% bold math
\usepackage{textcomp}
\usepackage[hidelinks]{hyperref}

\hypersetup{breaklinks=true}
\usepackage{epstopdf}
\def\ket#1{\mathinner{|{#1}\rangle}}

\begin{document}

\preprint{APS/123-QED}

\title{Observation of interference effects via four photon excitation of highly excited Rydberg states in thermal cesium vapor}% Force line breaks with \\

\author{Jorge M. Kondo}
\email{jorge.d.kondo@durham.ac.uk}
\author{{Nikola \v{S}ibali\'c}}%
\author{Alex Guttridge}
\author{Christopher G. Wade}
\author{Natalia R. De Melo}
\author{Charles S. Adams}
\author{Kevin J. Weatherill}
\affiliation{%
 Joint Quantum Centre (JQC) Durham-Newcastle, Department of Physics, Durham University, South Road, Durham, DH1 3LE, United Kingdom.
}%
\date{\today}% It is always \today, today,

\begin{abstract}

We report on the observation of Electromagnetically Induced Transparency (EIT) and Absorption (EIA) of highly excited Rydberg states in thermal Cs vapor using a four-step excitation scheme. The advantage of this 4-step scheme is that the final transition to the Rydberg state has a large dipole moment and one can achieve similar Rabi frequencies to two- or three-step excitation schemes using two orders of magnitude less laser power. This scheme enables new applications such as dephasing free Rydberg excitation. The observed lineshapes are in good agreement with simulations based on multilevel optical Bloch equations. 

\end{abstract}

\maketitle

%\section{Introduction}

The landscape of using atomic Rydberg states in an increasing variety of applications such as quantum computation \cite{RevModPhys.82.2313}, quantum simulation \cite{nature_quantum_sim}, quantum optics \cite{PhysRevLett.105.193603,Peyronel:2012fk}, elecric field sensors \cite{PhysRevA.84.023408,sadlacek_microwave,raithel_microwave}, single photon generation \cite{PhysRevLett.86.3534,Dudin18052012}, photon storage \cite{PhysRevLett.110.103001}, single-photon transistors \cite{PhysRevLett.113.053602,PhysRevLett.113.053601,PhysRevLett.107.133602}, etc., has led to a search for different routes for obtaining these highly excited states. In addition, Rydberg states are used to study a variety of physical processes including collective behaviour \cite{Schausz:2012uq,Weber:2015vn},  aggregate formation \cite{PhysRevLett.114.203002,PhysRevLett.112.013002}, and long-range molecule formation \cite{Bendkowsky:2009kx}, etc. Rydberg excitation can be achieved directly from the ground state \cite{PhysRevA.89.033416} but in most experiments the excitation is achieved using a three-level ladder system \cite{PhysRevLett.98.113003}. However, these one- and two-photon schemes presents some disadvantages. For example, the laser wavelengths often lie in the ultraviolet or blue range, necessitating costly and complicated second-harmonic generation systems. Also, the weak coupling between the ground-state  and the Rydberg state requires high power for laser excitation. Multi-photon schemes using lower-power near-infrared lasers provide a promising alternative and offer benefits such as easy light coupling to optical fibers, efficient excitation to Rydberg states, circumventing the accumulation of free charges due to photoelectric effects and background-free fluorescence detection of Rydberg states. Also, some off-axis beam geometries can be used to eliminate motional dephasing of Rydberg dark state polaritons \cite{PhysRevA.84.053409,PhysRevLett.108.030501} enabling efficient single spin-wave preparation.\par 
Electromagnetic Induced Transparency (EIT) has become the work-horse for coherent non-destructive detection of the properties of Rydberg states \cite{PhysRevLett.98.113003,reviewnonlinearRyd}. In previous work, EIT was demonstrated in a three-photon configuration for a four-level system in thermal Cs atoms \cite{Carr_12}. This excitation scheme enabled the observation of optical bistability and a non-equilibrium phase transition in thermal vapor \cite{PhysRevLett.111.113901}.
In this work we present the observation of interference effects in coherent four-step excitation of Rydberg states. The final step has a transition dipole moment eight-times larger compared to schemes using only two steps for the same Rydberg state (52D$_{3/2}$) or three steps for the same principal quantum number $\mathrm{n}$ with $\mathrm{l}=1$; therefore one may achieve the same driving Rabi frequency with 64 times less laser power. Following the ladder scheme $6\text{S}_{1/2}\rightarrow 6\text{P}_{3/2}\rightarrow 7\text{S}_{1/2}\rightarrow 8\text{P}_{1/2}\rightarrow 52\text{D}_{3/2}$ in cesium, the obtained spectra are in good agreement with a theoretical model based on the optical Bloch equations (OBEs). In addition we can observe the hyperfine splitting of the intermediate $8\text{P}_{1/2}$ state by three-photon EIT without the need for UV lasers, signal amplifiers, and long optical paths required in other works \cite{Cataliotti,PhysRevA.8.1661,long_cesium_cell}. This is due to the anomalously weak transition strength when exciting directly from the ground state. \par
\begin{figure}[htbp]
\centering\includegraphics[width=\linewidth]{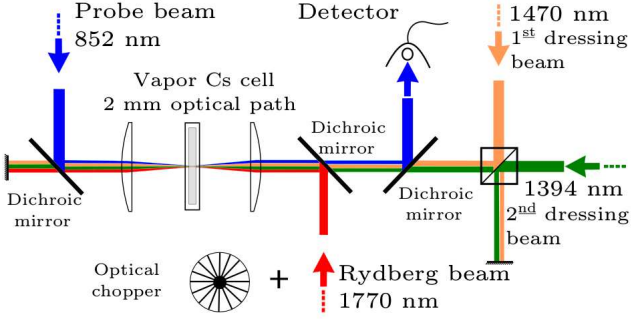}
\caption{Experimental setup: Four overlapping infrared beams (false color in diagram) are focused into a thermal cesium cell at $59\,^{\circ}$C. The probe transmission is measured at the photodiode detector.}
\label{figure1}
\end{figure}

The experimental setup consists of aligning four collinearly overlapped laser beams using dichroic mirrors and a polarising beam splitter as shown in Fig. \ref{figure1}. All beams are linearly polarized. The probe beam, which counterpropagates with respect to the direction of the other three beams, drives the first step of the process at 852 nm and is frequency stabilized by ground-state polarization spectroscopy \citep{PhysRevA.73.062509} to the transition $\ket{6\text{S}_{1/2},f=4} \rightarrow \ket{6\text{P}_{3/2},f'=5}$. The second step is driven by the first dressing beam at 1470 nm, stabilised to excited-state polarization spectroscopy \cite{carrESA} connecting states $\ket{6\text{P}_{3/2},f'=5} \rightarrow \ket{7\text{S}_{1/2},f''=4}$. The transition from $\ket{7\text{S}_{1/2},f''=4}$ to the hyperfine states of $\ket{8\text{P}_{1/2},f'''=\{3,4\}}$ marks the third step of the process and is driven by the second dressing beam at 1394 nm. This laser is not frequency stabilized but may be scanned or manually tuned by means of a low-noise current controller. Finally, the Rydberg beam at 1770 nm connects the hyperfine states of $8\text{P}_{1/2}$ to the Rydberg states $n\text{S}_{1/2}$ and $n\text{D}_{3/2}$ for $n\geq 52$.\par

All beams are focused to the center of a 2~mm long cesium vapor cell using 50~mm focal-length lenses. The beam waists are, in order from probe to Rydberg beam, 20, 49, 48 and 35~{\textmu}m respectively. A low-pass dichroic mirror is used to recover the probe beam after passing through the cell. The cell temperature is maintained at $59^{\circ}$C resulting in a number density $\mathcal{N}\approx1\times10^{12}\,\text{cm}^{-3}$ and $60\%$ absorption of the probe beam on resonance. Typical beam powers are $P_{852}=\,400$ nW, $P_{1470}=\,320$ {\textmu}W, $P_{1394}=\,500$ {\textmu}W and $P_{1770}=\,28.1$ mW with approximately Rabi frequencies of 44, 330, 156 and 30 MHz respectively. The second dressing beam is a commercial fiber-coupled laser (Qphotonics QDMLD-1392-10). The other lasers are commercial ECDL (Extended Cavity Diode Lasers) all with linewidths $<1$ MHz.

\begin{figure}[htbp]
\centering
\includegraphics[width=\linewidth]{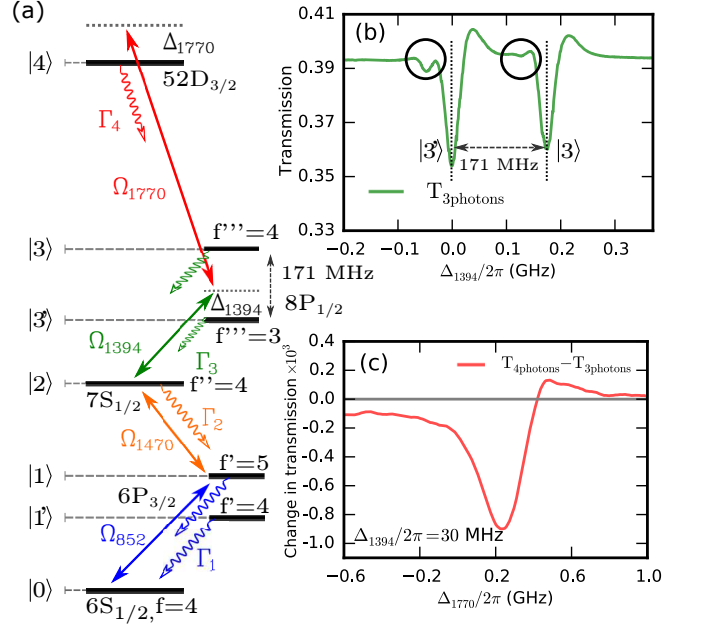}
\caption{(a) Probe, dressing and Rydberg beams couple ground state $\ket{0}$ to Rydberg state  $\ket{4}$. (b) Experimental data showing the hyperfine splitting of intermediate state $\{\ket{3'},\ket{3}\}=\ket{8\text{P}_{1/2},f'''=\{3,4\}}$ by three-photon interference, zero detuning correspond to transition $\ket{2} \rightarrow \ket{3'}$. (c) Measured probe-beam transmission when coupling to Rydberg state while scanning the Rydberg beam. The second dressing beam is detuned $\Delta_{1394}=30\pm3$ MHz from transition $\ket{2} \rightarrow \ket{3'}$.}
\label{figure2}
\end{figure}

%\section{Interference effect with three and four photons}
We measure the interference effects with three and four photons by monitoring the transmission of the probe beam. The levels coupled by the lasers are shown in Fig. \ref{figure2}(a). For simplification purposes we will rename the states as $\ket{0}=\ket{6\text{S}_{1/2},f=4}$, $\ket{1'}=\ket{6\text{P}_{3/2},f'=4}$, $\ket{1}=\ket{6\text{P}_{3/2},f'=5}$, $\ket{2}=\ket{7\text{S}_{1/2},f''=4}$, $\ket{3'}=\ket{8\text{P}_{1/2},f'''=3}$, $\ket{3}=\ket{8\text{P}_{1/2},f'''=4}$, $\ket{4}=\ket{52\text{D}_{3/2}}$. Without the Rydberg beam we were able to observe the hyperfine splitting of the intermediate $8\text{P}_{1/2}$ state using three-photon EIT \cite{Carr_12}. In this case the probe transmission is recorded while scanning the second dressing beam (1394 nm) over the $\ket{2} \rightarrow \{\ket{3'},\ket{3}\}$ transition while the first two-steps lasers are on resonance. The spectrum obtained  is shown in Fig. \ref{figure2}(b). The transmission line shape contains two absorptive features on three-photon resonance and is separated by the hyperfine splitting of $171$ MHz. This can be understood by considering a dressed-state picture where a pair of states are created by the first dressing beam $\ket{\Psi\pm}$ corresponding to the Autler-Townes (AT) splitting of state $\ket{1}$. The second dressing beam couples the split line to $8\text{P}_{1/2}$ state as follows: $\ket{0} \rightarrow \ket{\Psi\pm} \rightarrow \{\ket{3'},\ket{3}\}$. Here each path corresponds to an enhanced transmission on either side of the resonant three photon point and are detuned by $\Delta_{1394}/2\pi \approx \pm 50$ MHz given by the Autler-Townes effect. At zero detuning, paths through adjacent AT split states interfere constructively, leading to EIA on the-three photon resonance $\ket{0} \rightarrow \ket{1} \rightarrow \ket{2} \rightarrow \{\ket{3'},\ket{3}\}$. 

The transparency peak at the red side of the resonances are obscured by an absorption feature at $-45$ MHz from each hyperfine resonance highlighted in black circles in the Fig. \ref{figure2}(b). This is an effect of the sample atomic-velocity distribution. The Doppler shift, due to the group of atoms with velocities range around $213\,\text{m.s}^{-1}$, sets the probe beam in resonance to the hyperfine state $\ket{1'}$ detuned $-251$ MHz from the main resonance. The atoms within this velocity class fall in a Doppler-shifted three-photon resonance, which interferes constructively, destroying the transparency peak expected at the red side from both resonances. The overall size and widths of the features depends upon the Rabi frequencies of the two upper transitions since the Stark shifts partially cancel the Doppler shifts and more velocity classes contribute to the transmission signal.\par 
The four-photon transmission signal is obtained by turning on the Rydberg beam coupling the last step of the process while the second dressing beam is manually set to the desired detuning $\Delta_{1394}$ from transition $\ket{2} \rightarrow \ket{3'}$, whose  frequency is constantly monitored and recorded using a wavemeter (WS7 HighFinesse). This laser is stable during short-term scans and the maximum frequency drift at each value is 5~MHz. Fig. \ref{figure2}(c) shows the probe transmission while scanning the Rydberg laser frequency, the second dressing beam is fixed detuned to $\Delta_{1394}/2\pi=30\pm3$ MHz.
The maximum change in transmission of the probe beam due to the Rydberg beam is about $0.11\%$. Therefore, an optical chopper was used to modulate the Rydberg beam and a lock-in amplifier to demodulate the noise free signal. The overall behavior of the spectrum varies over different fixed detuning values $\Delta_{1394}$ showing both enhanced transparency and absorption.

%\section{4-photons Optical Bloch Equations (OBE)}

We can model the system by numerically solving OBEs or the so called Liouville–von Neumann equation \cite{PhysRevA.46.330} in the Lindblad form \cite{lindblad} given in Eq. (\ref{eq:1}). This is the equation of motion for the density matrix $\hat{\rho}$ describing the time evolution of the system,

\begin{equation}
\frac{d\hat{\rho}}{dt}=\frac{i}{\hbar}\left[\hat{\rho},\hat{\mathcal{H}}\right]+ \hat{\mathcal{L}}\left(\hat{\rho}\right)+\hat{\mathcal{L}}_{d}\left(\hat{\rho}\right).
\label{eq:1}
\end{equation}

The spontaneous decay rate among the states $\Gamma_{1}, \Gamma_{2}, \Gamma_{3}$, and $\Gamma_{4}$ are given by the following routes, respectively, $\{\ket{1'},\ket{1}\} \rightarrow \ket{0}, \ket{2} \rightarrow \{\ket{1'},\ket{1}\},\{\ket{3'},\ket{3}\} \rightarrow \ket{2}$ and $\ket{4} \rightarrow \{\ket{3'},\ket{3}\}$. These dissipative processes are taken in account by the Lindblad superoperator $\hat{\mathcal{L}}\left(\rho\right)$ acting to redistribute populations and dephase the coherences. Due to atomic motion, the atoms interact with the probe beam during a finite time. This transit time rate is accounted in the model as an average decay time of 74 ns from every excited state to the ground state $\ket{0}$. Extra dephasing effects due to finite laser linewidths $\gamma_{1,2,3,4}$, are phenomenologically added to Eq. (\ref{eq:1}) using the modified Lindblad operator in $\hat{\mathcal{L}}_{d}\left(\rho\right)$ and act only upon coherences.\par 
To reproduce the features of the experimental spectra we need to include the first excited hyperfine state $\ket{1'}$ in the model as it is crucial for the asymmetric behavior observed on the red side of the three photon EIT profile of the hyperfine state $8\text{P}_{1/2}$ Fig. \ref{figure2}(b). The total Hamiltonian representing the system in a rotating-wave approximation is the sum of the atom Hamiltonian $\hat{\mathcal{H}}_{atom}$ and light-coupling Hamiltonian $\hat{\mathcal{H}}_{light}$. The four-beams Rabi frequencies are given by $\Omega_{\{1,2,3,4\}}$ corresponding to the following wavelengths: 852, 1470, 1394 and 1770~nm, respectively, as displayed in the levels diagram in Fig. \ref{figure2}(a). The Doppler shifted detunings $\Delta_{\{1p,2p,3p,4p\}}$ due to co-propagating and counter-propagating beams are all accounted for in the model. The transmission-signal profile can be obtained by summing, from the steady state solution, the imaginary parts of matrix elements coupled by the probe laser given by $\rho_{01'}+\rho_{01}$ averaged over a Maxwell-Boltzmann velocity distribution. The numerical solution is realized using the approach discussed in \citep{Berman19911}.\par
\begin{figure}[t]
\centering
\includegraphics[width=\linewidth]{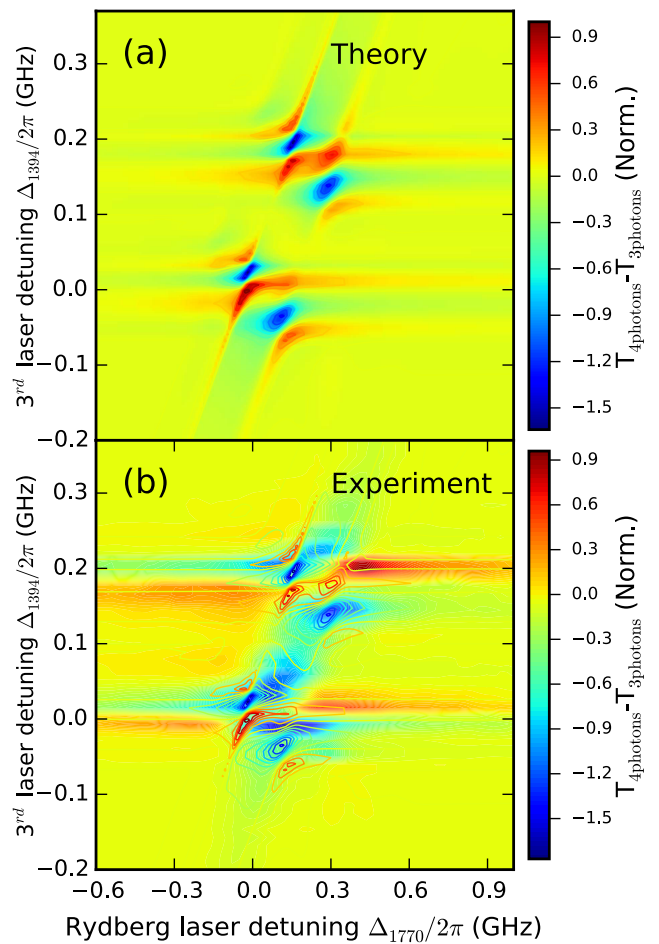}
\caption{(a) Normalized transmission from theoretical model depicting dependency to detunings $\Delta_{1394}$ and $\Delta_{1770}$ that couples states $7\text{S}_{1/2}\rightarrow 8\text{P}_{1/2}\rightarrow 52\text{D}_{3/2}$ respectively. (b) Comparison between numerical simulation and interpolated experimental data. Color lines mapped from numerical simulation are deployed on top of the experimental map to guide the eye.}
\label{figure3}
\end{figure}

 
The experimental three-step interference of the $8\text{P}_{1/2}$ hyperfine spectrum shown in Fig. \ref{figure2}(b) was used as the main comparison tool between the model and experiment. This three-photon interference is qualitatively reproduced by the model including the asymmetric line shapes. However, there is a mismatch between amplitudes of the features with the model predicting signals approximately twice as large as those observed in the experiment. We attribute this disagreement to experimental systematic errors such as imperfect beams polarizations, inhomogeneous Rabi frequencies, and decaying branching ratios to different states not included in the simple model. The complete four-photon numerical solution is presented in Fig. \ref{figure3}(a) and shows the dependency with the second dressing beam and Rydberg beam detunings $\Delta_{1394}$ and $\Delta_{1770}$, respectively. A background transmission signal related to the enhanced absorptions and transparencies of the intermediate state $8\text{P}_{1/2}$ is subtracted from the final spectra.\par

From the numerical solution in Fig. \ref{figure3}(a) we can see the formation of four distinct poles due to enhanced absorption (blue regions) symmetrically accompanied by transparencies (red regions) on either side of each resonant point. On four-photon resonance $\ket{0} \xrightarrow{\text{4-photons}} \ket{4}$ we observe enhanced transparency. The largest change in transmission is about $5\%$. The simulated transmission is normalized to the highest value. The quantum interference between all the excitation pathways lead to a complex change in the transmitted signal and result in appearance of "avoided crossings" like structures. Near both three-photon resonances, where the second dressing beam has detunings $\Delta_{1394}=0\;\text{MHz}\;\text{and}\;171\;\text{MHz}$, the transmission signal presents a transition from strict enhanced transparency to a strict absorptive feature in less than 5 MHz range, revealing a sensitive dependency to the second dressing beam detuning. \par
The four photon experiment was done by frequency stabilizing the probe beam (852 nm) and the first dressing laser (1470 nm) using ground-state saturation spectroscopy and excited-state saturation absorption, respectively. The second dressing beam (1394 nm) was manually tuned using a low noise currently controller to arbitrary detunings $\Delta_{1394}$ throughout the hyperfine states of $8\text{P}_{1/2}$ while its wavelength was constantly recorded. The Rydberg beam is frequency scanned by 2.6 GHz about the $52\text{D}_{3/2}$ Rydberg state. An optical chopper is used to modulate the Rydberg beam subtracting the background transmission signal due to the intermediate $8\text{P}_{1/2}$ profile while a lock-in amplifier demodulates the noise-free signal. The data was ordered using the recorded wavelength from the second dressing beam and an interpolation algorithm used to obtain the whole feature. We were able to observe change in probe transmission for higher Rydberg states up to n=80. However, the $52\text{D}_{3/2}$ state was chosen because it has the highest transition strength among the possible states accessible within the boundary limits of the laser diode emission.\par
The experimental results are shown in Fig. \ref{figure3}(b) where the absolute signal is normalized by the highest transmission value for comparison to theory. Color-contour lines from the numerical solution are displayed on top of the experimental data to guide the eye. Despite the very good agreement to the expected transmission signal, the data exhibits some distortion which can be explained by the lack of a frequency stabilization system on the second dressing beam causing drifting during the data acquisition. The result of this distortion is a smeared transmission map that washes out some of the narrow transmission features; although the main behaviour remains clear. The system is particularly sensitive at the four-photon resonance, due to a narrow absorptive feature, which is shown in Fig. \ref{figure2}(b), resulting in a fluctuation of the background signal that leads to a less clear transparency window at this position.

%\section{Summary}

In summary, we have presented the observation of interference effects in three- and four-photon ladder excitation scheme in a thermal cesium vapor.
Using a 2~mm optical-path length cell and modest laser powers up to only a few microwatts, we were able to obtain an EIT and EIA resolved spectrum of the hyperfine states of $8\text{P}_{1/2}$ whose measured splitting of $171\pm1$ MHz is in good agreement to previous measurement by different methods \cite{Cataliotti,PhysRevA.8.1661,long_cesium_cell}. We went on to demonstrate interference effects using four infrared lasers coupling the hyperfine ground state of cesium to a highly-excited Rydberg state. By solving the OBE numerically for a simple model averaged over different velocity classes we can reproduce the probe transmission map [Fig. \ref{figure3}(b)]. One slice is displayed in Fig. \ref{figure2}(c) for a fixed detuning of $\Delta_{1394}/2\pi=30\pm3$ MHz. To our knowledge this is the first observation of interference in a four-step cascade configuration in an atomic sample. Four-step cascade schemes have being proposed for other elements such as Rb \cite{PhysRevLett.108.030501}. 

This system has a variety of potential future applications such as the design of a fully fiber-coupled RF electrometer using simple diode laser systems \cite{sadlacek_microwave,raithel_microwave} and single photon nonlinear behavior \citep{Peyronel:2012fk} in the infrared domain preventing unwanted free photoelectric charge effects. The excitation system also permits Doppler-free configurations allowing the elimination of motional dephasing \citep{PhysRevA.84.053409}. This results in narrower interferences linewidths and presents alternative excitation schemes to highly excited Rydberg states used in the search for Rydberg blockade in micrometre-sized atomic-vapor cells \cite{microcell}. The data presented in this Letter are available \cite{dataset}.\par
\vspace{10pt}
%\section{ACKNOWLEDGEMENT}

\noindent
We would like to acknowledge the financial support given by Durham University, The Federal Brazilian Agency of Research (CNPq), The Royal Society "(Grant No. PI 120001)" and EPSRC "(Grant No.EP/M014398/1)".

\end{document}